\documentclass[sigconf,natbib=true]{acmart}

\AtBeginDocument{%
  \providecommand\BibTeX{{%
    \normalfont B\kern-0.5em{\scshape i\kern-0.25em b}\kern-0.8em\TeX}}}

\copyrightyear{2024} 
\acmYear{2024} 
\setcopyright{acmlicensed}
\acmConference[SIGIR '24]{Proceedings of the 47th International ACM SIGIR Conference on Research and Development in Information Retrieval}{July 14--18, 2024}{Washington, DC, USA}
\acmBooktitle{Proceedings of the 47th International ACM SIGIR Conference on Research and Development in Information Retrieval (SIGIR '24), July 14--18, 2024, Washington, DC, USA}
\acmDOI{10.1145/3626772.3661343}
\acmISBN{979-8-4007-0431-4/24/07}

\usepackage[utf8]{inputenc}
\usepackage{graphicx}
\usepackage{amsmath}
\usepackage{algorithm}
\usepackage{algorithmic}
\usepackage{CJKutf8}
\usepackage{subcaption}
\usepackage{balance}
\usepackage{multirow}
\usepackage[skip=2.5pt]{caption}
\usepackage{caption}

\newcommand\blfootnote[1]{%
  \begingroup
  \renewcommand\thefootnote{}\footnote{#1}%
  \addtocounter{footnote}{-1}%
  \endgroup
}

\begin{document}

\setlength{\abovedisplayskip}{2.5pt}
\setlength{\belowdisplayskip}{2.5pt}

\title{Optimizing E-commerce Search: Toward a Generalizable and Rank-Consistent Pre-Ranking Model}


\author{Enqiang Xu}
\author{Yiming Qiu$\,^{*}$}
\email {xuenqiang@jd.com}
\email {qiuyiming3@jd.com}
\affiliation{%
  \institution{JD.com}
  \country{Beijing, China}
}
\author{Junyang Bai}
\author{Ping Zhang}
\author{Dadong Miao}
\author{Songlin Wang}

\affiliation{%
  \institution{JD.com}
  \country{Beijing, China}
}
\author{Guoyu Tang}
\author{Lin Liu}
\author{MingMing Li$\,^{*}$}
\email {limingming65@jd.com}
\affiliation{%
  \institution{JD.com}
  \country{Beijing, China}
}

\renewcommand{\shortauthors}{Enqiang Xu et al.}
\ccsdesc[500]{Information systems}
\ccsdesc[500]{Information systems~Novelty in information retrieval}
\ccsdesc[500]{Information systems~Information retrieval}
\ccsdesc[500]{Information systems~Retrieval models and ranking}

\begin{abstract}
In large e-commerce platforms, search systems are typically composed of a series of modules, including recall, pre-ranking, and ranking phases. The pre-ranking phase, serving as a lightweight module, is crucial for filtering out the bulk of products in advance for the downstream ranking module. Industrial efforts on optimizing the pre-ranking model have predominantly focused on enhancing ranking consistency, model structure, and generalization towards long-tail items.
Beyond these optimizations, meeting the system performance requirements presents a significant challenge. Contrasting with existing industry works, we propose a novel method: a $\mathbf{G}$eneralizable and $\mathbf{RA}$nk-$\mathbf{C}$onsist$\mathbf{E}$nt Pre-Ranking Model ($\mathbf{GRACE}$), which achieves: 1) Ranking consistency by introducing multiple binary classification tasks that predict whether a product is within the top-k results as estimated by the ranking model, which facilitates the addition of learning objectives on common point-wise ranking models; 2) Generalizability through contrastive learning of representation for all products by pre-training on a subset of ranking product embeddings; 3) Ease of implementation in feature construction and online deployment.
Our extensive experiments demonstrate significant improvements in both offline metrics and online A/B test: a 0.75\% increase in AUC and a 1.28\% increase in CVR.

\blfootnote{$^*\,$ Corresponding author}
\end{abstract}

\keywords{Embedding index, neural network, information retrieval}

\maketitle

\section{Introduction}
\label{sec:introduction}

Large-scale E-commerce platforms potentially engage in the sale of billions of items, catering to a user base in the hundreds of millions. 
To enhance user experience and conversion efficiency, the online search system is employed with a cascading architecture, mainly including recall and ranking.
The ranking stage as the downstream component directly influences the efficiency of item sorting. Several superior ranking models have been identified in industrial research, such as MMoE~\cite{ma2018modeling}, PLE~\cite{tang2020progressive}, ESMM~\cite{ma2018entire}, 
DeepFM~\cite{guo2017deepfm},
DIN~\cite{zhou2018deep}, MIMN~\cite{pi2019practice},
SDIM~\cite{zhao2020dimension}
, and SIM~\cite{tang2020progressive}, with a focus on feature engineering, behavioral sequence modeling, and objective function optimization.
However, as the scale of products within the search system grows, there is an increasing demand for managing the time complexity of the sorting module. To this end, the industry often divides the sorting process into pre-ranking and ranking models, operating in tandem to balance the trade-off between system processing time and the accuracy of sorting.
To some extent, the effectiveness of pre-ranking sets the upper limit for the ranking, thereby affecting the performance of the entire search system. 

There is an increasing scholarly interest in optimizing pre-ranking models. Commonly, dual-tower models or simple interaction models are employed as the foundational model. Representative dual-tower models like DSSM~\cite{huang2013learning} calculate the similarity between queries and documents for scoring and ranking to achieve optimal efficiency. Models like PFD~\cite{xu2020privileged} introduce interacted features into the learning of pre-ranking models through the distillation of representations from ranking models. Single-tower models exemplified by COLD~\cite{wang2020cold}, which incorporate SE (Squeeze-and-Excitation) blocks in the feature selection process, balance effectiveness, and computational resources, with experimental data showing optimized single-tower models achieving better results without significantly increasing prediction time.   

Beyond the optimization of ranking models themselves, it is essential to address the distinct challenges that pre-ranking models face compared to ranking models. We identify several key challenges: 1) \textbf{Consistency with ranking}: As previously discussed, the goal of pre-ranking models is to filter out items early in the system to save computational resources for ranking. Achieving closer ranking model performance with fewer model parameters is the core intent behind the design of pre-ranking modules. In addition to adopting approximate objective functions, features, and model structures, the industry has proposed direct approaches to model consistency. RankFlow~\cite{qin2022rankflow} introduces a two-stage model, where the first stage involves self-learning using online feedback labels, and the second stage incorporates downstream task scoring, although learning scores of ranking model is challenging. COPR~\cite{zhao2023copr} suggests using relative positioning in ranking as pair-wise supervision signals in pre-ranking systems for advertisements, adding weight factors to penalize incorrect ordering of top pairs. While this method circumvents the issue of learning ranking scores, it requires data transformation to construct pair relationships within the same session. 2) \textbf{Generalization}: It is widely accepted that simply placing ranking models in the position of pre-ranking does not yield positive results, indicating that fitting pre-ranking models to the ranking are not optimal. Pre-ranking models predict a larger array of long-tail items, whereas ranking models aim to sort within the sequence output by pre-ranking. Thus, pre-ranking models require improved item representation and long-tail generalization capabilities. Meta-Embedding~\cite{pan2019warm} proposes a supervised framework to optimize both cold-start and warm-up advertisement representations, but this method is limited by only using attribute features of items to generate cold-start embeddings and is not as straightforward due to its two-stage training process. MWUF~\cite{zhu2021learning} introduces optimizations and simplifies the meta-learning process, yet this approach is still not end-to-end and does not fully leverage attribute features for cold-start items.

\begin{figure}[t]
	\includegraphics[scale=0.42]{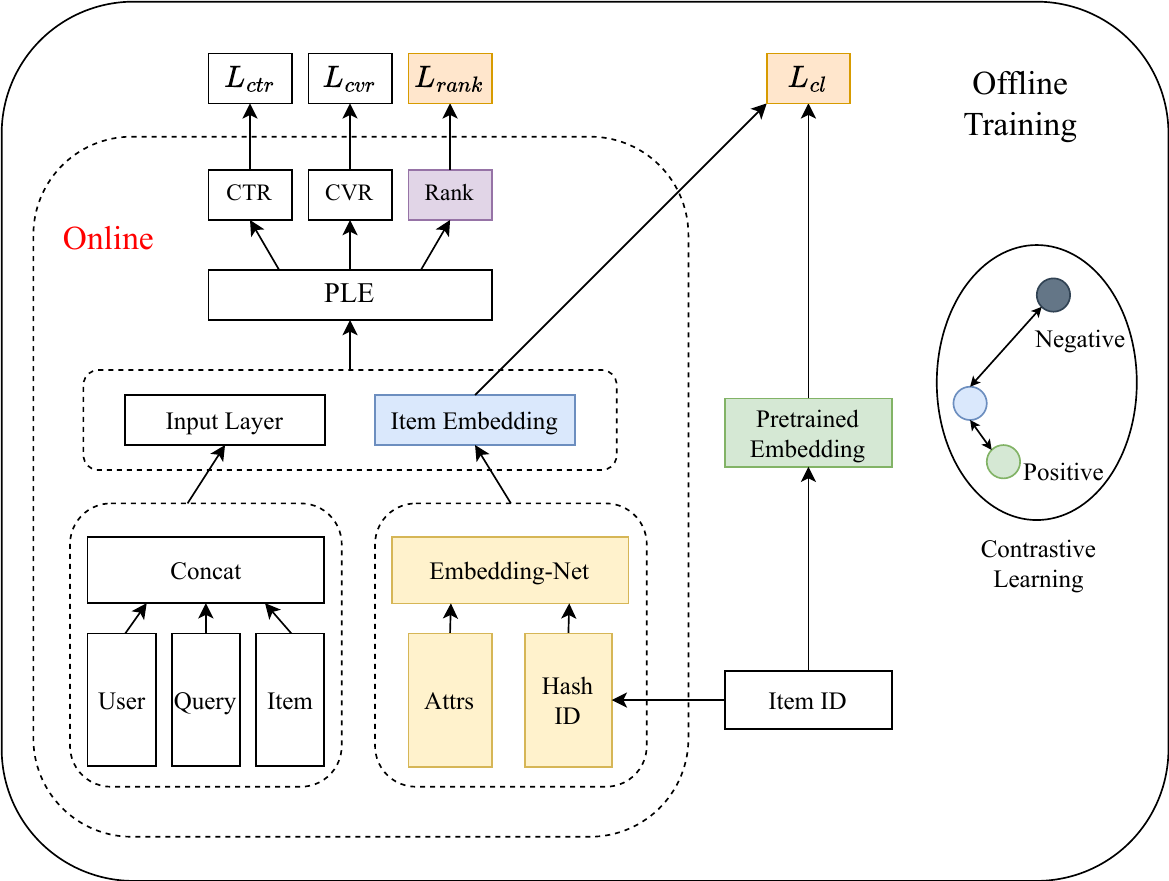}
	\caption{The proposed framework, where "Attrs" refers to attribute information such as brand, shop, and category. }
	\label{fig:model_structure}
 \vspace{-4mm}
\end{figure}

To alleviate the above problems, we propose an innovative approach: a $\mathbf{G}$eneralizable and $\mathbf{RA}$nk-$\mathbf{C}$onsist$\mathbf{E}$nt Pre-Ranking Model ($\mathbf{GRACE}$). This method is designed to enhance performance by addressing two critical aspects: rank consistency and generalizability. Figure~\ref{fig:model_structure} illustrates the overall architecture of the model.

Firstly, for rank consistency, GRACE innovatively incorporates a consistency task within the pre-ranking framework, which is built upon a multi-task baseline model~\cite{tang2020progressive}. Specifically, it leverages position information from ranking training data (online logs) and learns whether a given sample belongs to the top k positions as determined by the ranking model. This is achieved by introducing an additional loss function, which requires no modification to the existing training data or process, thereby simplifying implementation.

Secondly, for generalizability, our model begins by hashing item IDs to generate hash ID embeddings, which are initialized and looked up at the scale of millions, and then fuses these with attributes embeddings (initialized and looked up similarly, for attributes such as brand, shop, category, etc.) to initialize item representations through concatenation. To enhance the accuracy of these embeddings, we introduce supervision signals derived from embeddings generated by a pre-trained Graph Neural Network (GNN) and employ Info-NCE~\cite{oord2018representation} loss for end-to-end learning. Our model, with reduced parameter size, achieves comparable accuracy to models trained directly with embeddings of millions of items. Furthermore, unlike methods relying on pre-trained embeddings of a limited number of items, GRACE supports the generalization across an unlimited range of long-tail items.

The optimizations integrated into GRACE do not impose significant storage or computational burdens, whether in offline training or online deployment, making it an elegant and straightforward adaptation for industrial model enhancements. Empirical validation through offline experiments demonstrated a 0.749\% improvement in the area under the ROC curve (AUC~\cite{xu2020privileged}). In online A/B testing, GRACE achieved a 1.26\% increase in conversion rate (CVR) and a 1.62\% rise in gross merchandise value (GMV), with even more pronounced improvements for long-tail items, where we observed a 2.89\% boost in CVR and a 10.37\% increase in GMV. These results underscore the effectiveness of GRACE in addressing the nuanced demands of pre-ranking models in e-commerce search engines.

\section{Method}
\label{sec:method}
\subsection{Task Definition}
Given a user $U$ entering a query $Q$ into the search bar, a set of pre-ranking candidate products $S = \{\text{item}_1, \text{item}_2, \ldots, \text{item}_M\}$ is generated, where $M$ denotes the number of pre-ranking candidate products. The model computes predictions $\hat{y} = F(U, Q, S)$, from which the top $N$ products are selected and returned to the following ranking phase, with $N \ll M$. In our training dataset, we utilize user feedback from online logs as labels, where the click-through rate (CTR) records whether a user clicks and the conversion rate (CVR) records whether a user converts.

\subsection{Base Model}
The baseline is the Progressive Layered Extraction~\cite{tang2020progressive} (PLE) multi-task model which addresses the trade-off issue between multiple tasks by employing a combination of shared and task-specific expert networks. 
We input the aforementioned variables $U$, $Q$, and $S$ into the model to predict the click-through rate (CTR) and conversion rate (CVR), as shown in the following equations:

\begin{equation}
    \hat{y}^{ctr},  \hat{y}^{cvr} = Sigmoid\left ( PLE(U, Q, S) \right ) \in [0,1] 
\end{equation}

We employ the cross-entropy loss to compute the loss for each task, where $y^{ctr}$ and $y^{cvr}$ represent the click label and conversion label respectively, denoted as:

\begin{equation}
    \mathcal{L}_{ctr} = -\frac{1}{N} \sum_{i=1}^{N} y^{ctr}_{i} log(\hat{y}^{ctr}_{i}) +(1 - y^{ctr}_{i})log(1 - \hat{y}^{ctr}_{i})
\end{equation}

\begin{equation}
    \mathcal{L}_{cvr} = -\frac{1}{N} \sum_{i=1}^{N} y^{cvr}_{i} log(\hat{y}^{cvr}_{i}) +(1 - y^{cvr}_{i})log(1 - \hat{y}^{cvr}_{i})
\end{equation}

The total loss of base model could be formulated as:

\begin{equation}
    \mathcal{L}_{feedback} = \mathcal{L}_{ctr} + \mathcal{L}_{cvr}
\end{equation}

\subsection{Rank Consistency Loss}
For the pre-ranking candidate items, we get their positions after ranking, where $pos^{rank}_{i}$ represents the position of $item_i$ after scoring in the ranking model.

\begin{equation}
 \left \{ (item_{1}, pos^{rank}_{1}), ..., (item_{N}, pos^{rank}_{N}) \right \}   
\end{equation}

Depending on whether the product's position falls within the top $k$, we establish a binary classification task to determine its membership in the top-$k$ subset.

\begin{equation}
    y^{k}  = \begin{cases}
        1, & pos^{rank} \le  k\\
        0, & pos^{rank} >  k
\end{cases}
\end{equation}

In the following, we extend the initial pre-ranking task by additionally predicting the objective of rank consistency.

\begin{equation}
    \hat{y}^{ctr},  \hat{y}^{cvr},  \hat{y}^{rank} = Sigmoid\left ( PLE(U, Q, S) \right ) \in [0,1] 
\end{equation}

Then, we could get the loss of consistency, denoted as:

\begin{equation}
    \mathcal{L}_{k} = -\frac{1}{N} \sum_{i=1}^{N} y^{k}_{i} log(\hat{y}^{rank}_{i}) +(1 - y^{k}_{i})log(1 - \hat{y}^{rank}_{i})
\end{equation}

To further refine the accuracy of the top-$k$ retrieval task, we employ multiple values of $k$ and aggregate the corresponding losses to compute the overall ranking loss $L_{\text{rank}}$, where $w_k$ is the hyper-parameter.

\begin{equation}
    \mathcal{L}_{rank} = \sum_{k\in K} w_{k} \mathcal{L}_{k}
\end{equation}

\subsection{Generalization Loss}
We incorporate an auxiliary contrastive learning task to enforce the representation of items to align closely with pre-trained embeddings. Let ${P}$ denote the set of pre-trained items, $\phi$ represent the embedding network, and $\psi$ the pre-trained embeddings. Given a batch size ${N}$ and an inner product \( \langle \cdot, \cdot \rangle \), with $\tau$ as the temperature coefficient, the contrastive learning loss is computed only for positive instances with existing pre-trained embeddings. Specifically, the aforementioned InfoNCE~\cite{oord2018representation} loss encourages a reduction in the distance between the representations of items and their corresponding pre-trained embeddings, while promoting an increase in the distance between item embeddings across different product categories. 

\begin{equation}
    \mathcal{L}_{cl} = -\sum_{ x_{i} \in P } log\frac{exp\left ( \left \langle \phi \left ( x_{i}  \right ),\psi \left ( x_{i}  \right )   \right \rangle /\tau  \right ) }{\sum_{ j = 0}^{N} exp \left ( \left \langle \phi \left ( x_{i}  \right ),\phi \left ( x_{j}  \right )   \right \rangle /\tau  \right ) }
\end{equation}

\subsection{Total Loss}
The final optimization objective of the model is defined below, where ${\lambda}_1$ and ${\lambda}_2$ are hyper-parameters.

\begin{equation}
    \mathcal{L} = \mathcal{L}_{feedback} + \lambda _{1}\mathcal{L}_{rank} + \lambda _{2}\mathcal{L}_{cl}
\end{equation}

\section{Experiments}
\label{sec:experiments}

\subsection{Setup}
\subsubsection{Dataset}
Our training dataset is compiled from 15 days of search logs, amassing approximately one billion records. Each record is annotated with multiple binary encoded labels indicating whether an item is clicked and ordered (click label and conversion label). As to negative instances, they comprise two categories: one sourced from user feedback, such as items displayed but not clicked, and the other from random sampling of items not displayed to enrich the pool of negative samples. For model validation, we utilized data from the entire 16th day to evaluate the model's performance.

\subsubsection{Metrics}
\begin{itemize}
\item \underline{AUC}: Area Under Curve, is a performance measurement for classification problems, quantifying the ability of a model to distinguish between classes~\cite{xu2020privileged}.
\item \underline{GMV}: Gross Merchandise Value, a metric that measures the total value of sales for merchandise sold through a particular marketplace over a specific time period~\cite{qiu2021query,qiu2022pre}. 
\item \underline{CVR}: Conversion Rate, widely adopted indicators within E-commerce platforms \cite{adaptive,multi-obj}.
\item \underline{Recall@k}: measures the consistency between pre-ranking and ranking results by calculating the intersection of the top-k items from both the pre-ranking and ranking models, divided by the top-k items from the ranking model.

\begin{equation}
    recall @k = \frac{\left | \left \{ topk_{pre-ranking}  \right \}  \cap \left \{ topk_{ranking}  \right \} \right |  }{ \left | \left \{ topk_{ranking}  \right \}   \right |  } 
\end{equation}

\end{itemize}

\begin{table}
    \begin{minipage}{.22\textwidth}
        \centering
        \caption{Offline metrics.}
        \begin{tabular}{c|c}
        \toprule
           &  AUC  \\ \midrule
        PLE  &  0.7713  \\
        MMOE &  0.7705  \\  
        Distillation &  0.7727  \\  \midrule
        GRACE & $\mathbf{0.7788}$ \\
        \bottomrule
        \end{tabular}
        \label{tab:offline}
    \end{minipage}
    \begin{minipage}{.22\textwidth}
        \centering
        \caption{Ablation study.}
        \begin{tabular}{c|c}
        \toprule
           &  AUC  \\ \midrule
        w/o consistency &  0.7753  \\  
        w/o generalization &  0.7747  \\  
        pre-trained item id & 0.7775 \\
        w/o hash id  &  0.7775 \\ \midrule
        GRACE  &  $\mathbf{0.7788}$  \\ 
        \bottomrule
        \end{tabular}
        \label{tab:ablation}
    \end{minipage}
    \vspace{-.6cm}
\end{table}

\subsubsection{Baseline models}
For the sake of a fair comparison, we select the most representative works from the industry, such as PLE and MMoE, against which to benchmark our model. Our model is an optimization directly based on the PLE model. Additionally, to validate our consistency approach, we add a model that employs direct distillation of ranking scores for comparison. Specifically:
\begin{itemize}
\item \underline{PLE}: Represents our baseline model deployed online.
\item \underline{MMoE}: Stands for the Multi-gate Mixture-of-Experts, a commonly used multi-task model in the industry.
\item \underline{Distillation}: Represents the model utilizing the distillation of ranking scores.
\end{itemize}

\subsubsection{Hyperparameter}
As to the top-k value for the consistency loss in our method, we empirically chose 10, 30, 50, and 100 as parameters for our experiments, which are determined by factors such as the browsing depth of users on different e-commerce platforms or the number of products displayed per page on the front-end interface. On our platform, each page displays 10 items, and the average browsing depth of users is 30. In the loss function, ${\lambda}_1$ and ${\lambda}_2$ are both set to 0.1. As for the model parameters: the batch size is 1024, the learning rate is 0.05, the optimizer is Adagrad, and the number of epochs is 20. For the hash ID, the embedding dimension is 32, and the vocabulary size is 3 million.

\subsection{Comparison with baseline models}
From Table~\ref{tab:offline}, it is evident that our model shows a significant improvement over the PLE model, with the AUC absolute value increasing by 0.749\%. For reference, when compared to another baseline model, MMoE, the increase is even more substantial. This demonstrates that the improvement margin of our model far exceeds the gap between MMoE and PLE. Furthermore, in comparison with the mainstream industry method of distillation, our results show that distillation can bring about a 0.14\% increase in AUC. However, the gap remains significant compared to our model, validating our earlier discussion that directly learning from ranking scores is not necessarily the optimal solution.

\begin{table}[t]
\centering
\caption{Consistency analysis about Recall@k.}
\begin{tabular}{c|ccccc}
\toprule
   &  R@3 & R@10 & R@20 & R@50 & R@100  \\ \midrule
PLE   & 0.5316 & 0.4995 & 0.4908 & 0.5067 & 0.5422 \\ 
GRACE   & $\mathbf{0.5549}$  & $\mathbf{0.5216}$ & $\mathbf{0.5110}$ & $\mathbf{0.5256}$ & $\mathbf{0.5612}$ \\  \bottomrule
\end{tabular}
\label{tab:consistency}
\vspace{-2mm}

\end{table}

\begin{table}[t]
\centering
\caption{Online A/B test relative improvement. p-value is obtained by t-test, small p-value means statistically significant.}
\begin{tabular}{c|c|c}
\toprule
   &  CVR &   GMV  \\ \midrule
Overall   &  +1.28\% &   +1.62\%  \\
p-value & 0.0008& 0.0040\\ \midrule
Long-tail   & +2.89\%   & +10.37\%  \\  
p-value & -& - \\ \bottomrule

\end{tabular}
\label{tab:online}
\vspace{-4mm}
\end{table}

\subsection{Ablation Study}
In our model, we have introduced multiple enhancements over PLE, as detailed in the comprehensive ablation experiments presented in Table~\ref{tab:ablation}. This allows us to analyze the sources of AUC gains. The abbreviation "w/o" in the table stands for "without."

\subsubsection{Generalization module}
The ablation results in table~\ref{tab:ablation} demonstrate that the generalization module contributes to a 0.41\% increase in AUC, sufficiently proving its effectiveness. Furthermore, we conducted experiments from two perspectives to further validate the significance of the generalization module:

\begin{itemize}
\item \underline{w/o hash id}: the straightforward way was to generate generalized representations using product attribute features. However, it became apparent that many product attributes are repetitive, leading to duplicate representations, such as those for products from the same brand or shop, which could potentially conflict and reduce the accuracy of the representations. This duplication posed a significant challenge in learning pretrain embeddings. To address this, we introduced hashed ID as an initialization, which, after a lookup, yielded generalized ID embeddings with a smaller parameter footprint. We posited that hashed ID, combined with product attributes, could serve as unique representations. Upon removing the hashed ID from our model, we observed a 0.13\% decrease in AUC, confirming the effectiveness of hashing.
\item \underline{pre-trained item id}: We removed the generalization module from our model and replaced it with pre-trained embeddings, which, despite their large parameter size (with a vocabulary size of 30 million and an embedding dimension of 32), still fall short of covering all products. This substitution significantly increased the storage burden for the pre-ranking model. Our findings indicate that GRACE slightly outperforms the one with pre-trained embeddings, demonstrating that we have successfully learned a comparable effect with fewer parameters. Moreover, our model provides more accurate representations for long-tail products, leading to a further increase in AUC.
\end{itemize}

\subsubsection{Rank consistency module}

 The ablation study on the consistency module presented in Table~\ref{tab:ablation} shows that the experiments lacking this module exhibit an AUC that is 0.35\% lower than that of the GRACE model, conclusively demonstrating the efficacy of the consistency module.

\subsubsection{Rank consistency analysis}
 Analysis of Table~\ref{tab:consistency} reveals a marked increase in recall rates for both pre-ranking and ranking sets, with enhancements ranging from 3.50\% to 4.43\%. Notably, the largest gains were recorded in recall@3 and recall@10, metrics of heightened importance on our platform, which displays a maximum of 10 products per page on the front-end.

\subsection{Online A/B Test}
On an e-commerce platform with tens of millions of daily active users, we conducted a 7-day A/B test. Variant A deployed the baseline PLE model, while Variant B deployed GRACE model, with each variant receiving 10\% of the traffic. From table~\ref{tab:online}, we can observe a cumulative increase of 1.28\% in conversion rate (CVR) and 1.62\% in gross merchandise volume (GMV). By examining the long-tail traffic, which accounts for approximately 10\% of the total traffic, we confirmed the significant generalization effects of the proposed model, achieving a 2.89\% increase in CVR and a 10.87\% increase in GMV within this segment. GRACE was deployed in a production environment at the end of 2022.
\section{Conclusion}
\label{sec:conclusion}
In our work, we introduce GRACE, a novel pre-ranking model that addresses the dual challenges of rank consistency and generalizability. By innovatively incorporating position information from ranking, GRACE achieves rank consistency without requiring modifications to the training data or process. For generalizability, we fuse hash ID embeddings with attribute-based representations, enabling effective scaling to billions of items without significant storage.
Compared with related works, GRACE demonstrates competitive results on both offline studies and online A/B test.

\bibliographystyle{ACM-Reference-Format}
\balance

\newpage
\bibliography{references}

\section*{Company Portrait}
JD.com, Inc., also known as Jingdong, is a Chinese e-commerce company headquartered in Beijing. It is one of the two massive B2C online retailers in China by transaction volume and revenue, a member of the Fortune Global 500. When classified as a tech company, it is the largest in China by revenue and 7th in the world in 2021.

\section*{Presenter profiles}
\noindent\textbf{Enqiang Xu} is a researcher in the Department of Search and Recommendation at JD.com Beijing. 
He received his master degree in School of Mathematical Sciences, Peking University.
His research focuses on information retrieval and natural language processing.

\noindent\textbf{Yiming Qiu} is a researcher in the Department of Search and Recommendation at JD.com Beijing. He received his master degree in School of Computing and Information System, the University of Melbourne. His research focuses on information retrieval and natural language processing.

\noindent\textbf{Mingming Li} is a researcher in the Department of Search and Recommendation at JD.com Beijing. 
He received his doctor degree in Institute of Automation, Chinese Academy of Sciences. 
His research focuses on information retrieval and natural language processing.

\end{document}